%% file: conf_lp2005.tex
\documentclass[11pt]{article}
\usepackage{graphicx}

\newcommand{\BABARPubYear}    {05}

\newcommand{\BABARConfNumber} {10}
\newcommand{\SLACPubNumber} {11323}

\input{include/babarsym.tex}
\usepackage{cite}
\input{include/phys}
\input{include/results}

\long\def\inst#1{\par\nobreak\kern 4pt\nobreak
    {\it #1}\par\vskip 10pt plus 3pt minus 3pt}

\def\up{\rule[0mm]{0mm}{3ex}}
\def\down{\rule[-1.5ex]{0mm}{3ex}}
\def\both{\rule[-1.5ex]{0mm}{4.5ex}}
\usepackage{amsmath,amssymb}

\usepackage[small]{caption}

\begin{document}
{\pagestyle{empty}

\begin{flushright}
 \babar-CONF-\BABARPubYear/\BABARConfNumber \\
 SLAC-PUB-\SLACPubNumber \\
 July 2005 \\
\end{flushright}

\par\vskip 4cm

\begin{center}
\Large \bf A Study of Production and Decays of \OmegacZ\ Baryons at \babar\
\end{center}
\bigskip

\begin{center}
\large The \babar\ Collaboration\\
\mbox{ }\\
\today
\end{center}
\bigskip \bigskip

\begin{center}
\large \bf Abstract
\end{center}
Production and decay of \OmegacZ baryons is studied with
$\sim230\invfb$ of data recorded with the \babar\ detector at the
PEP-II \epem asymmetric-energy storage ring at SLAC. The
\OmegacZ\ is reconstructed through its decays  into \FOmegapi, 
\FOmegapipipi, and \FXiKpipi\ final states. 

The invariant mass spectra are presented and the signal yields are
extracted. Ratios of branching fractions are measured relative to the
\Omegapi\ mode
\begin{align*}
  \frac{{\cal B}(\XiKpipi)}{{\cal B}(\Omegapi)}     
         &= \RXiKpipi \pm \RstatXiKpipi\mathrm{(stat.)} 
         \pm \RsysXiKpipi\mathrm{(syst.)},\\
  \frac{{\cal B}(\Omegapipipi)}{{\cal B}(\Omegapi)} 
        &< \LOmegapipipi\qquad (90\%\mathrm{CL}).
\end{align*}

The momentum spectrum (not corrected for efficiency) of \OmegacZ\
baryons is extracted from decays into \FOmegapi, establishing the
first observation of \OmegacZ\ production from $B$ decays.

\vfill
\begin{center}
Contributed to the 
XXII$^{\rm nd}$ International Symposium on Lepton and Photon Interactions at High~Energies, 6/30 --- 7/5/2005, Uppsala, Sweden
\end{center}

\vspace{1.0cm}
\begin{center}
{\em Stanford Linear Accelerator Center, Stanford University, 
Stanford, CA 94309} \\ \vspace{0.1cm}\hrule\vspace{0.1cm}
Work supported in part by Department of Energy contract DE-AC03-76SF00515.
\end{center}

\newpage
} 

\input include/authors_conf05010.tex

\section{INTRODUCTION}
\label{sec:Introduction}
\renewcommand{\thefootnote}{\fnsymbol{footnote}}
The \OmegacZ\ ($css$) is a ($J^P=\frac{1}{2}^+$)\footnote[3]{The
  quantum numbers have not been measured, but are assigned in accord
  with the quark model.} ground state baryon
with a mass of $m_{\OmegacZ}=(2697.5\pm2.6)\mevcc$ and a lifetime 
of $\tau_{\OmegacZ}=(69\pm12)\fs$~\cite{bib:PDG2004}.
Since the first evidence for \OmegacZ\ production and decay in
1984~\cite{Biagi:1984mu} in the decay mode \XiKpipi, the \OmegacZ
baryon has been seen by a number of different
experiments~\cite{Albrecht:1992xa, Frabetti:1992bm,
  Frabetti:1994dp, Ahmed:2000fd, Adamovich:1995pf, Link:2003nq} in
various decay modes, with strong evidence reported
in the decays \Omegapi, \XiKpipi, and \SigmaKKpi. To
date, only one $5\sigma$ observation has been reported by a single
experiment, combining two modes~\cite{Link:2003nq}. No observation of
a single exclusive \OmegacZ\ decay mode at the $5\sigma$ level has
been reported and only a few of its decay modes have been
observed. The measurements of the ratios of branching fractions still
have large uncertainties and the production mechanisms of the
\OmegacZ\ remain largely unexplored.   
\renewcommand{\thefootnote}{\arabic{footnote}}

The large amount of data collected at the $B$ factories allows a more
detailed analysis of the properties of the \OmegacZ. In this paper, a
study of the \OmegacZ through the decay modes\footnote{Simultaneous
  treatment of the charge conjugate mode is always implied throughout
  the note.} \FOmegapi, \FOmegapipipi, and \FXiKpipi\ is
described. 

The invariant mass spectra are presented and the ratios of branching
fractions relative to the \FOmegapi\ decay mode are calculated. To
study the production mechanism, the \OmegacZ\ momentum spectrum (not
corrected for efficiency) in the \epem\ rest frame is presented.    

\section{THE \babar\ DETECTOR AND DATASET}
\label{sec:babar}
The \babar\ detector operating at the PEP-II \epem asymmetric-energy
storage ring at SLAC consists of a tracking system for the detection
of charged particles, a detector of internally reflected Cherenkov
light (DIRC), an electromagnetic calorimeter (EMC), and an instrumented flux
return (IFR). The tracking system, contained in a 1.5-T magnetic field
provided by a superconducting solenoidal coil, includes a 5-layer,
double-sided silicon vertex tracker (SVT) and a 40-layer drift chamber
(DCH). The EMC consists of 6580 CsI(Tl) crystals. Information from the
DIRC and the energy loss information from the SVT and the DCH are used
for charged  particle identification (PID). The IFR is segmented and
instrumented with resistive plate chambers. 
The \babar\ detector is described in detail 
elsewhere~\cite{bib:babar}.     

This analysis is based on data taken at the \FourS\ resonance
and $\sim40\mevcc$ below, referred to as on-peak and
off-peak data, respectively. The integrated luminosity of the data used
corresponds to $\sim 225\invfb$ in each of the \OmegacZ\ decay modes
involving an \Omegam, and $\sim 230\invfb$ in the \Xim\ decay mode.

To minimize the selection bias, all selection criteria are optimized
on Monte Carlo simulated event samples at least as large as the data
sample. Samples of $\epem\to q\bar q$ ($q\in\{u,\,d,\,s,\,c\}$) events
are used to study the background. Monte Carlo samples are
used to study the signal properties and to evaluate the selection
efficiencies. The \OmegacZ\ signal samples are generated with
JETSET~\cite{bib:Jetset}, assuming uniform phase space for the
\OmegacZ\ decays.

\section{EVENT SELECTION}
\label{sec:Analysis}
The \Omegam\ (\Xim) is reconstructed in its \LambdaZ\Km\
(\LambdaZ\pim) final state. Candidates for \LambdaZ\ decays are
reconstructed in their $p\pim$ final state. The branching
fractions~\cite{bib:PDG2004} of the 
intermediate hyperons are ${\cal
  B}(\Omegam\to\LambdaZ\Km)=(67.8\pm0.7)\%$, ${\cal
  B}(\Xim\to\LambdaZ\pim)\sim100\%$ , and ${\cal B}(\LambdaZ\to p\pim)=(63.9\pm0.5)\%$.  

\subsection{The Hyperon Selection}
All hyperons in this analysis (\LambdaZ, \Xim, and \Omegam) are
long-lived particles with a typical decay length of several cm in
\babar. Each hyperon is identified by reconstructing its decay
vertex, which is required to be clearly displaced from that of the
parent particle. In the selection of each of the intermediate
hyperons, only candidates with an invariant mass within $3\sigma$ of
the central value are selected, where $\sigma$ denotes the invariant
mass resolution. These candidates are then subjected to a kinematic
fit, constraining the mass of the candidate to its nominal value. 

Candidates of \LambdaZ\ baryons are formed from a pair of tracks of
opposite charge, where the positively charged track must satisfy the
PID requirements for a proton. Each \LambdaZ\ candidate is combined
with a \pim\ (\Km) to form a \Xim\ (\Omegam) candidate. The \Km\
candidate track from the \Omegam\ decay must satisfy the PID
requirements for kaons.    

Candidates for \OmegacZ\ decays for each final state are formed by
combining the reconstructed \Omegam\ and \Xim\ baryons
with the required number of mesons (\pim,\,\pip, or
\Km). Throughout this document, mesons that are the
direct daughters of the \OmegacZ\ candidate are referred
to as \lq primary mesons\rq.  

\subsection{The \OmegacZ\ Candidate Selection}
The kinematics of the signal decays and the backgrounds are different for
each of the \OmegacZ\ decay channels considered in this
analysis. The $Q$-values, which are a measure of the
available kinetic energy of the decay products, are $882\mev$,
$603\mev$, and $595\mev$ for the \FOmegapi, \FOmegapipipi, and
\FXiKpipi\ decay modes, respectively. Therefore, the selection criteria
are optimized for each mode separately. Particle identification is
required for each of the primary mesons. The selection criteria
described below are then applied to the \OmegacZ\ candidates in order
to suppress backgrounds. In addition, a common minimum $p^\ast$ of
$2.8\gevc$ is required for candidates used in the measurement of the
ratios of branching fractions, where $p^\ast$ is  the \OmegacZ\
momentum in the \epem\ rest frame. This is above the kinematic limit
for \OmegacZ\ production from $B$ decays ($p^\ast=2.02\gevc$).  

\begin{itemize}
\item \Omegapi:\par
  The transverse flight length of the \Omegam\ candidate in the
  $xy$-plane calculated with respect to the event
  vertex is required to exceed $2\mm$. The signed\footnote{The signed
  flight length is  defined as the dot product of the displacement
  vector $(\vec r_{\LambdaZ} - \vec r_{\Omega})$ of the \LambdaZ\ and the
  momentum vector of the \LambdaZ, where $\vec r$ denotes the 3D
  position of the vertex. }  
  flight length of the \LambdaZ, measured from the \Omegam\ decay
  point, must exceed $1.5\mm$. For the primary pion, a
  minimum momentum of $200\mevc$ in the laboratory frame is required. A
  minimum of $0.1\%$ is required for the $\chi^2$-probability of the
  kinematic fit for each intermediate hyperon.  
\item \Omegapipipi:\par
  The transverse flight length of the \Omegam\ candidate in the
  $xy$-plane calculated with respect to the event
  vertex is required to exceed $2.5\mm$. The signed flight length of
  the \LambdaZ, measured from the \Omegam\ decay point, must exceed
  $2\mm$. The vector sum of the momenta of the  $\pip \pim \pip$
  system in the lab-frame is required to exceed $650\mevc$. A minimum
  of $0.1\%$ is required for the $\chi^2$-probability of the kinematic
  fit for each intermediate hyperon.   

\item \XiKpipi:\par
  The minimum requirement on the flight length of the \Xim\ is
  $4.5\mm$ and the signed flight length of the \LambdaZ\ with respect
  to the \Xim\ decay vertex is required to be larger than zero. A
  minimum $\chi^2$ probability of $10^{-4}$ is required for the
  kinematic fit of the full decay chain.  
\end{itemize}
\clearpage

\section{THE INVARIANT MASS SPECTRA}
\begin{figure}[!b]
  \centering\small
  \includegraphics[width=.49\textwidth]{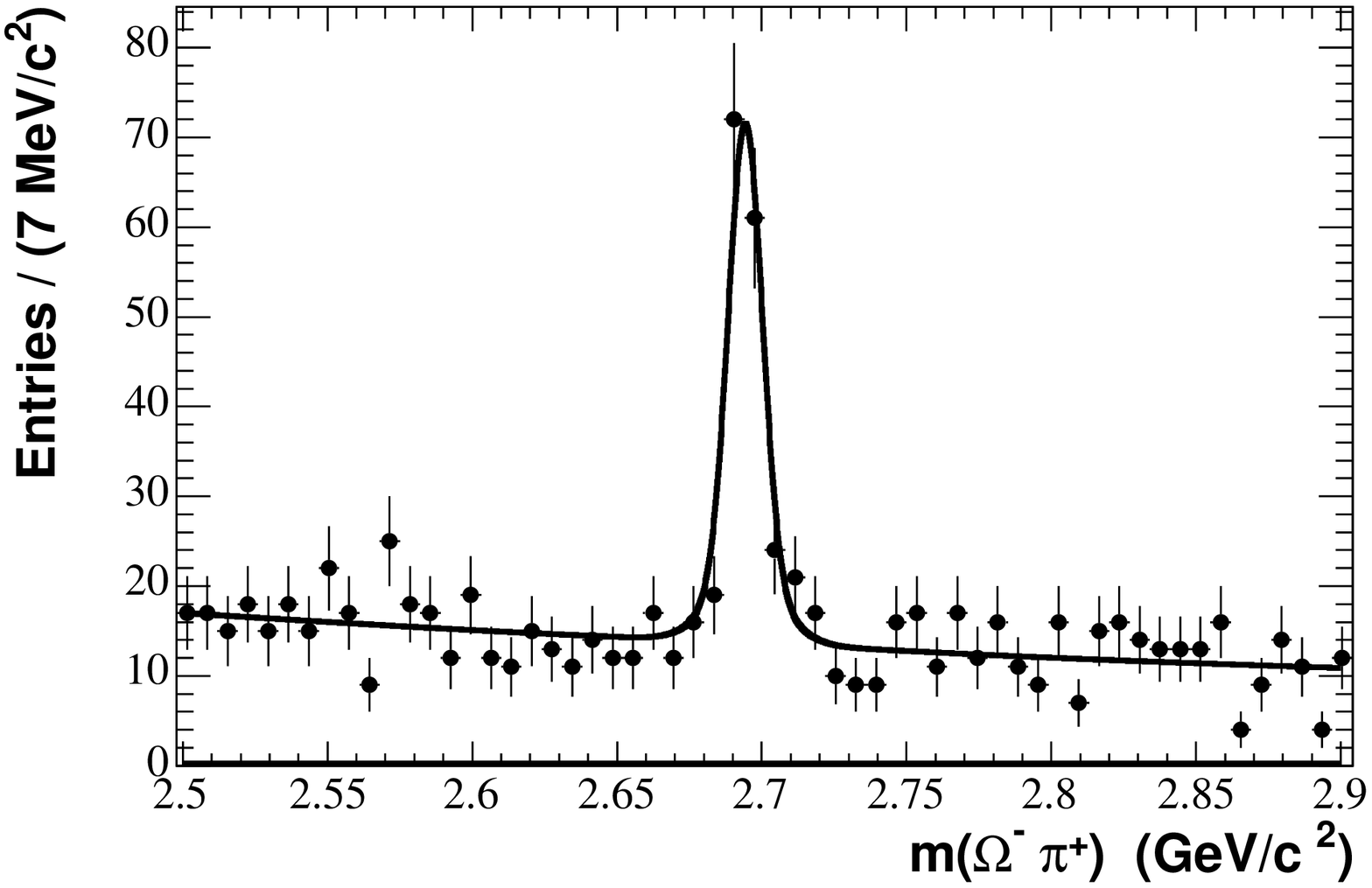}
  \includegraphics[width=.49\textwidth]{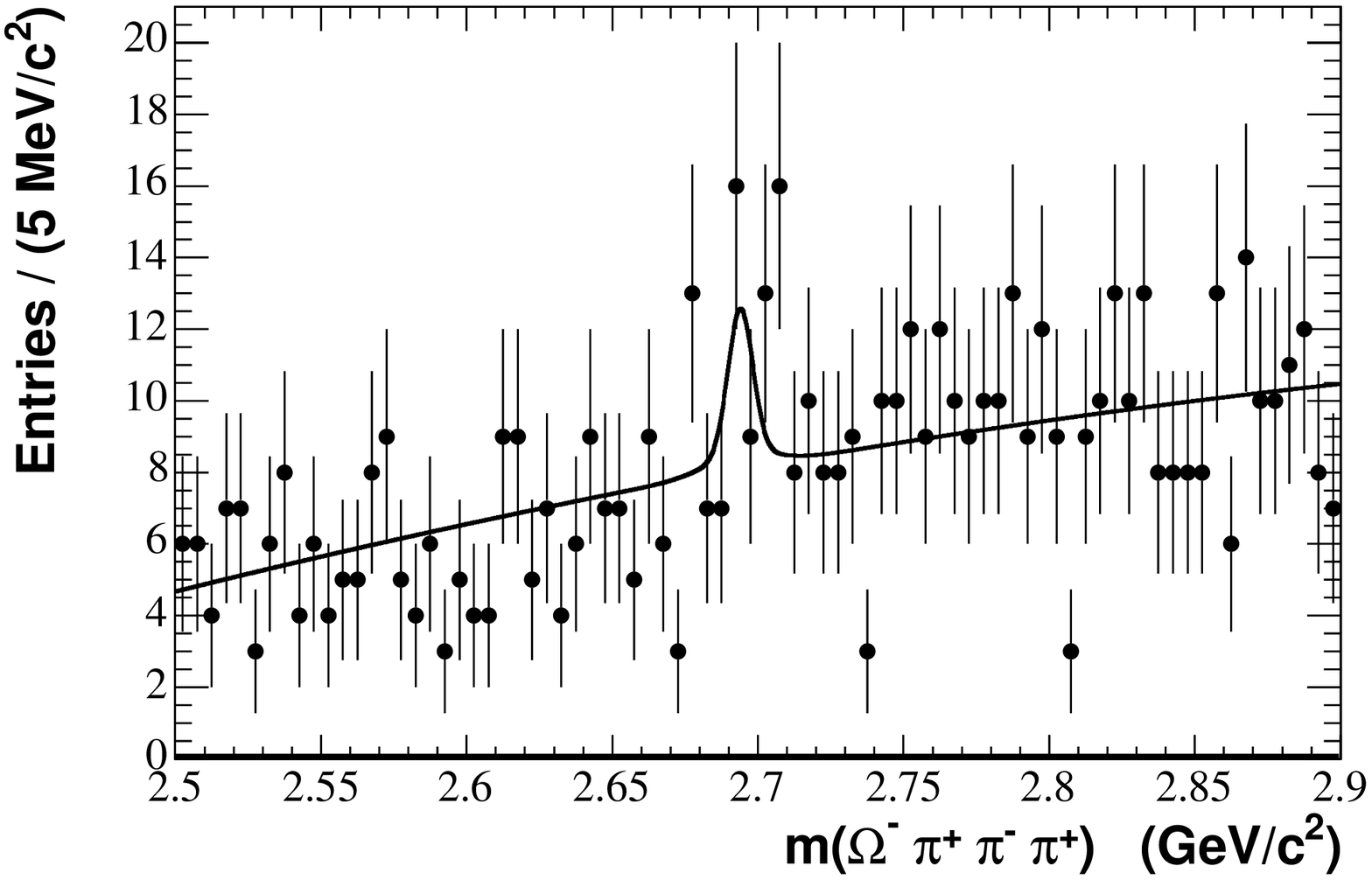}
  \includegraphics[width=.49\textwidth]{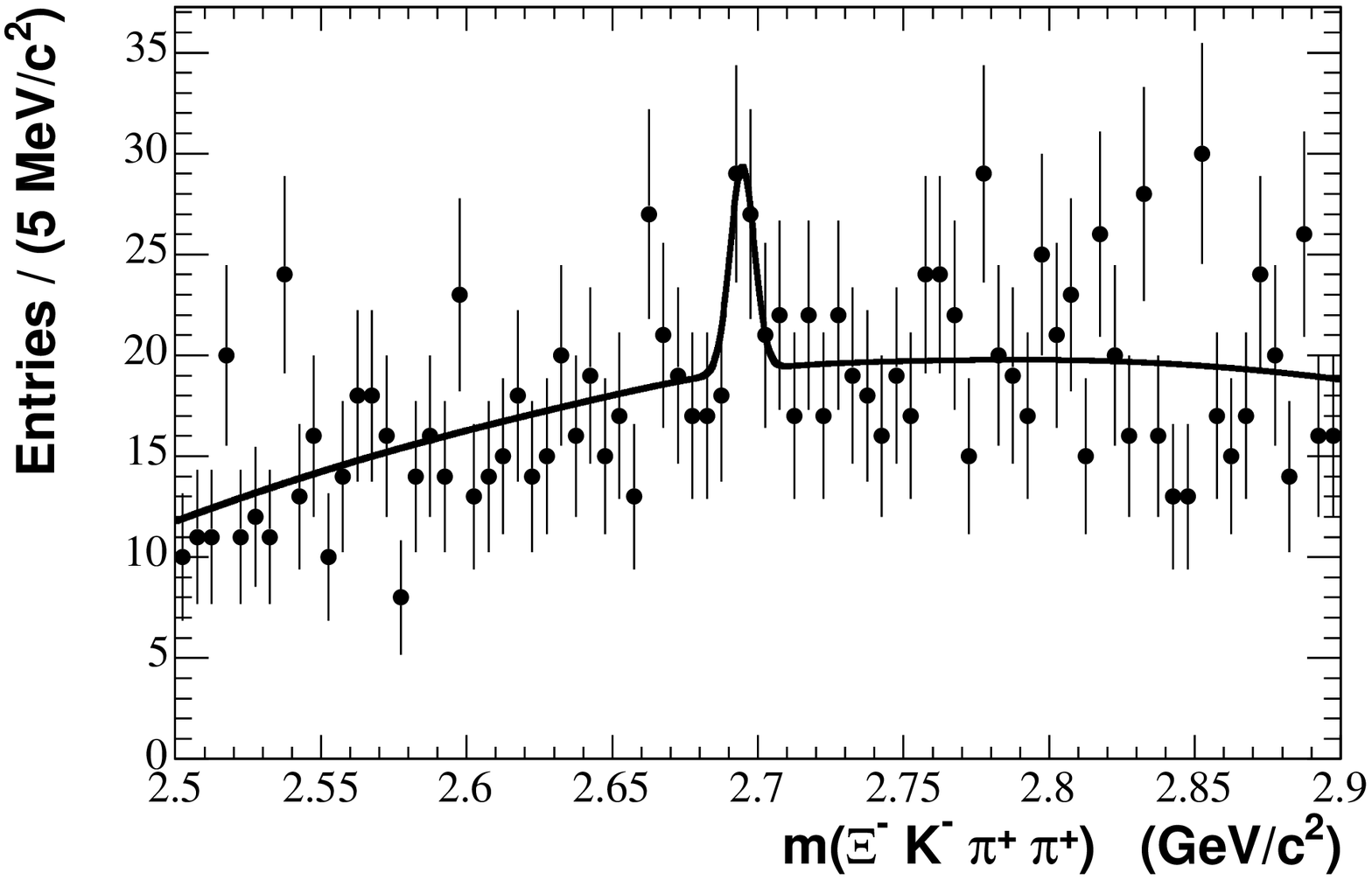}\\
  \begin{picture}(0.,0.)
    \put(-190,300){\bf (a)}
    \put( 185,300){\bf (b)}
    \put( 80, 50){\bf (c)}
    \put(-80,300){\normalsize\babar}\put(-80,290){preliminary}
    \put( 50,300){\normalsize\babar}\put( 50,290){preliminary}
    \put(-65,145){\normalsize\babar}\put(-65,135){preliminary}
  \end{picture}
  \caption{Invariant mass spectra for \OmegacZ\ decays into (a)
    \FOmegapi, (b) \FOmegapipipi, and (c) \FXiKpipi
    final states with $p^\ast>2.8\gevc$. The dots are the data. The
    binning is chosen according to the Half-Width-Half-Maximum of the
    signal peak obtained from signal Monte Carlo. The result of the
    fit is overlaid} 
  \label{fig:DataInvMass}
\end{figure}
The signal yield for each decay mode is extracted from an
unbinned maximum likelihood fit to the invariant mass spectrum of the
\OmegacZ\ candidates. In the
fit, the signal lineshape is described by a sum of three
Gaussians with a common mean, and the background is described by a
first order (\FOmegapi\ mode) or second order (all other modes)
polynomial. The widths and relative contributions of each Gaussian are
fixed to the values obtained from samples of simulated \OmegacZ\
baryons produced in $c\bar c$ continuum and decayed in the considered
mode. 
The mean mass, the yield of the signal peak, and the parameters in
the polynomial for the description of the background are left free
in the fit to the data. In the fit to the \FOmegapipipi\ mode, due to
the small signal, the mean of the signal
lineshape is fixed to the weighted mean\footnote{This is not necessarily
the true mass of the \OmegacZ\ baryon. No studies of systematic biases
due to energy loss corrections of the decay products or due
to the uncertainties on the energy and momentum scale are carried
out.}  
($\mu=2694.6\mevcc$) obtained from the fits to the other two decay modes.    

The invariant mass spectra are displayed in
Figures~\ref{fig:DataInvMass}(a-c) for the \FOmegapi,
\FOmegapipipi, and \FXiKpipi\ modes. The results of the
fits to the data, the selection efficiencies\footnote{All selection
  efficiencies quoted are for \OmegacZ\ decays with
  $p^\ast>2.8\gevcc$.}, and the $\chi^2$ 
probability for each fit are summarized in
Table~\ref{tab:DataInvMass}. In addition, the significance ${\cal S}$
for the signal, defined as $\sqrt{2\Delta\log{\cal L}}$, calculated
from the difference in the log-likelihood ($\log{\cal L}$) for a fit
with and without a signal lineshape, is included. 

\begin{table}[!t]
  \centering\small
  \caption{Results from the fits to the invariant mass spectra in
    data. The yields (not corrected for efficiency) are given
    for each mode individually, as well as the selection
    efficiencies. In addition the $\chi^2$ probability for the fit,
    calculated in the mass window $2.6<m<2.8\gevcc$, and the
    significance ${\cal S}$ of the signal is quoted for the
    \FOmegapipipi and the \FXiKpipi\ modes.}\vspace*{1ex}
  \label{tab:DataInvMass}
  \begin{tabular}{l|cccc}\hline\hline
    \both Decay Mode    & Signal Yield  & Efficiency (\%) & $\mathrm{prob}(\chi^2)$ & ${\cal S}$\\\hline 
    \up   \FOmegapi     & $\RYOmegapi    \pm\RYstatOmegapi    $ & $\EffOmegapi     \pm\DEffOmegapi    $ & $0.54$ & $17.8$ \\
          \FOmegapipipi & $\RYOmegapipipi\pm\RYstatOmegapipipi$ & $\EffOmegapipipi \pm\DEffOmegapipipi$ & $0.73$ & $2.4$ \\
    \both \FXiKpipi     & $\RYXiKpipi    \pm\RYstatXiKpipi    $ & $\EffXiKpipi     \pm\DEffXiKpipi    $ & $0.85$ & $3.4$ \\\hline\hline
  \end{tabular}  
\end{table}

\section{PRODUCTION MECHANISM FOR \OmegacZ\ BARYONS}
Insight into the production mechanism for \OmegacZ\ baryons is
obtained from the $p^\ast$ spectrum. For this study, the
\Omegapi\ decay mode, which has the largest signal yield of all modes in
this analysis, is used.
The signal yield as a function of $p^\ast$ is measured
up to $4.4\gevc$ in eleven intervals, each $400\mevc$ wide. For each
interval, the \FOmegapi\ invariant mass spectrum is fit with the
lineshape for the signal fixed to that obtained from
the full signal Monte Carlo sample. A polynomial is used to  describe
the background. No significant variation in the central value of the signal
peak is observed as a function of $p^\ast$. It is therefore fixed to
the common mean for all intervals in the fit.

\begin{figure}[!t]
  \centering\small
  \includegraphics[width=.49\textwidth]{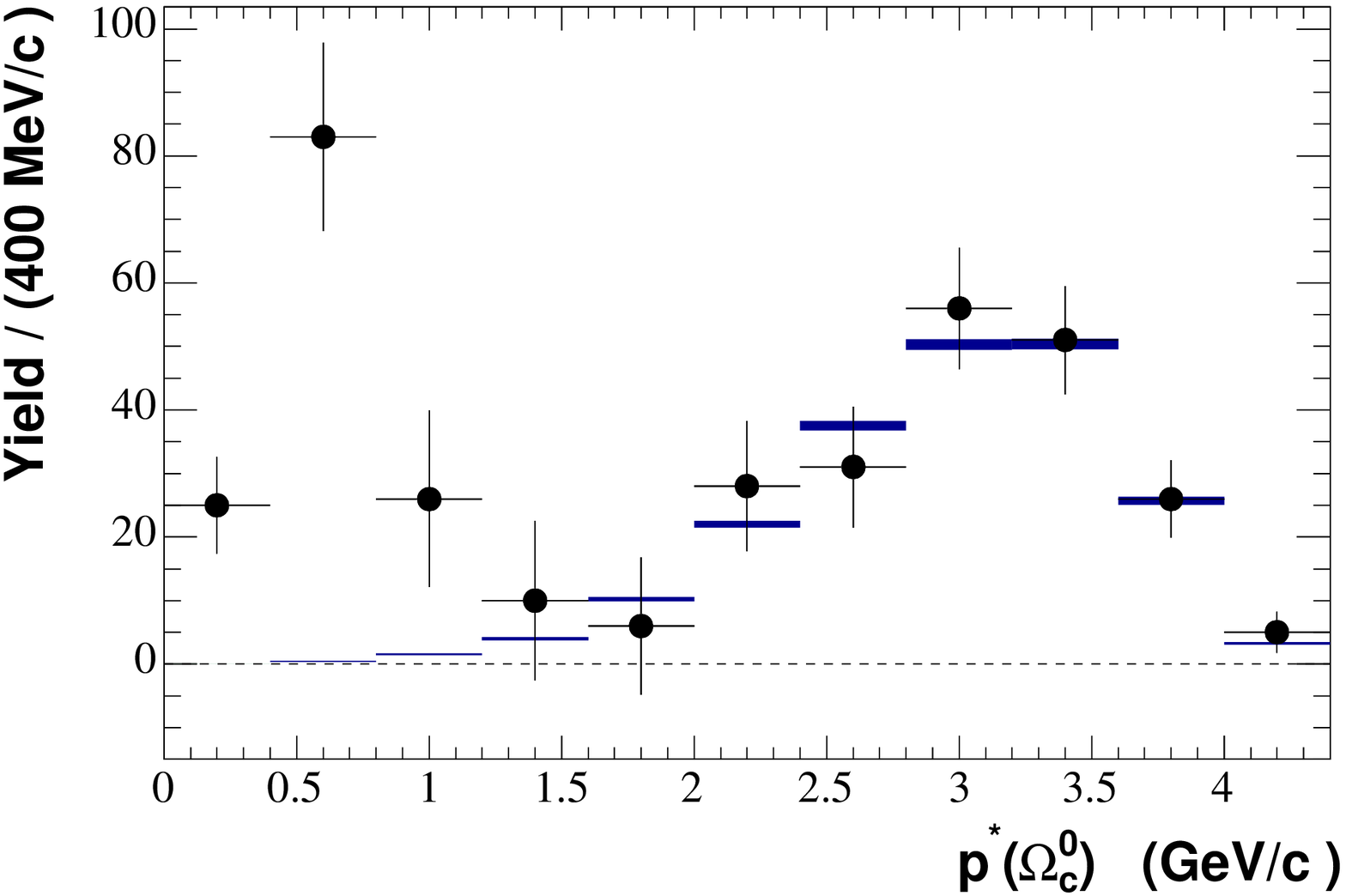}
  \includegraphics[width=.49\textwidth]{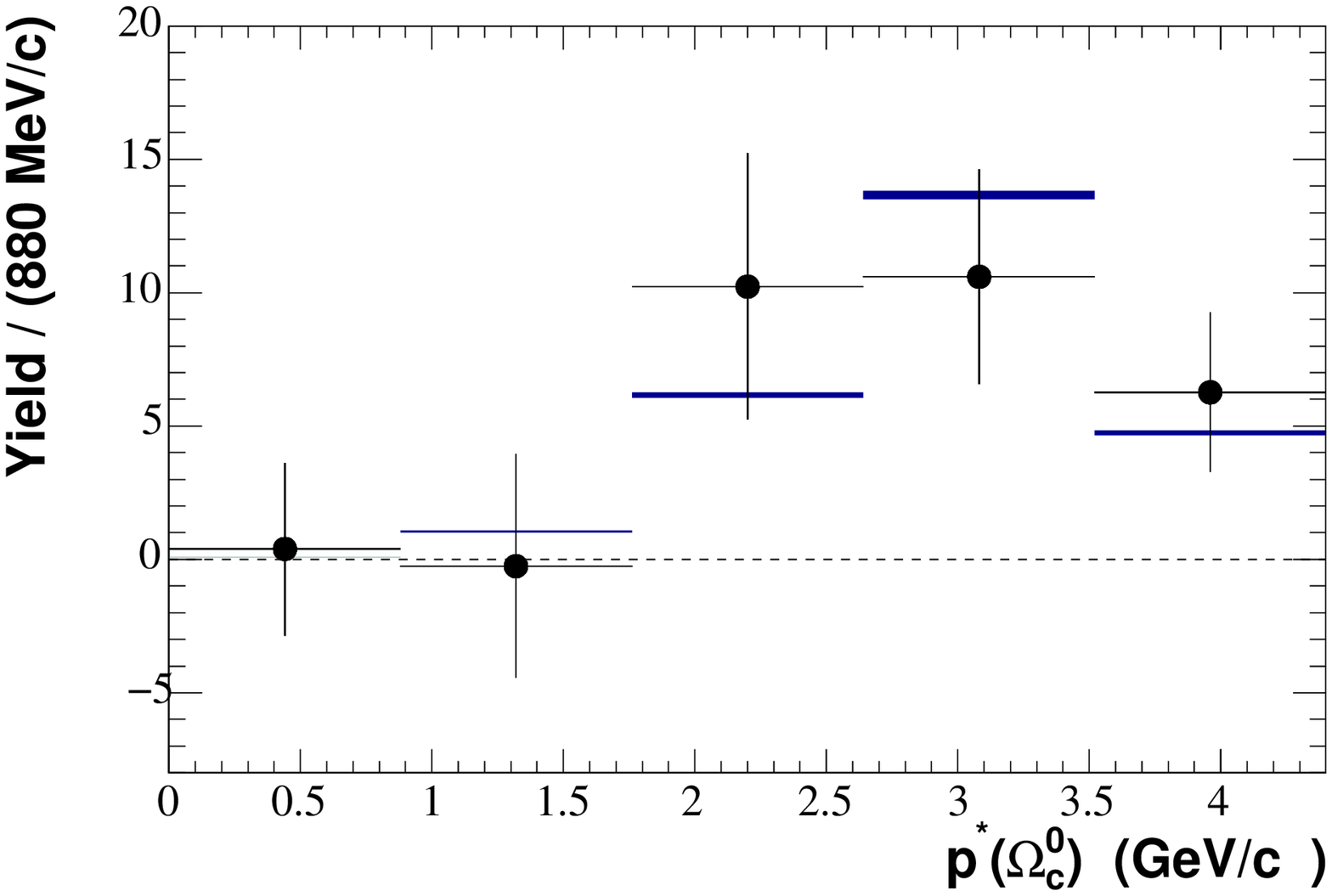}
  \begin{picture}(0.,0.)
    \put(-390,130){\normalsize\babar}
    \put(-390,120){preliminary}
    \put(-390,110){uncorrected}
    \put(-190,130){\normalsize\babar}
    \put(-190,120){preliminary}
    \put(-190,110){uncorrected}
    \put(-270,130){\bf(a)}
    \put( -40,130){\bf(b)}
    \end{picture}
  \caption{The signal yield as a function of the $p^\ast$ of \OmegacZ\
    candidates (a) from the combined on-peak and off-peak data sets and
    (b) from off-peak data only. The dots are the data and the
    vertical error bars correspond to the statistical uncertainty
    only. The solid horizontal bars correspond to the predicted
    distribution for \OmegacZ\ production from $c\bar c$ continuum
    signal Monte Carlo. The thickness of the bars correspond to the
    statistical  uncertainty of the Monte Carlo sample. No correction
    for selection efficiency is applied to any of the distributions
    shown. The distributions are normalized to the same area for (a)
    $p^\ast>2.5\gevc$ and (b) the full range. In (a), clear evidence
    for \OmegacZ\ production from $B$ decays is visible at low
    $p^\ast$. It is absent in off-peak data (b), collected below the
    $B\overline{B}$ production threshold.}     
  \label{fig:pAst}
\end{figure}

The measured signal yields obtained from the combined on-peak and off
peak data sets can be
compared with those from the off-peak data set, displayed in
Figures~\ref{fig:pAst}(a) and \ref{fig:pAst}(b), respectively. The
dots represent the data and the error bars correspond to the
statistical uncertainty only. The solid horizontal bars correspond to
the predicted spectrum for \OmegacZ\ production from $c\bar c$
continuum Monte Carlo; the thickness of the bars correspond to the
statistical uncertainty. No correction for selection 
efficiency is applied to either distribution. The corresponding
distribution from signal Monte Carlo is normalized such that its
integral corresponds to that in data for  $p^\ast>2.5\gevc$ in
Figure~\ref{fig:pAst}(a), and for the full $p^\ast$ range in
Figure~\ref{fig:pAst}(b). 

A clear two-peak structure is evident in the distribution from the
combined on-peak and off-peak data sets. The peak at high $p^\ast$ is
consistent with \OmegacZ\ production as predicted from continuum
signal Monte Carlo and the off-peak data. The $p^\ast$ spectra from
data and Monte Carlo show good agreement within the experimental
uncertainties in this region. The peak in the $p^\ast$ region below
$2.02\gevc$ provides clear first evidence for \OmegacZ\ production
from $B$ decays. This interpretation is substantiated by the absence
of the corresponding peak in the spectrum extracted from off-peak data
only, taken below the $B\bar B$ threshold.   

\section{PHYSICS RESULTS}
\label{sec:Systematics}
\subsection{Ratios of Branching Fractions}
The yields for \OmegacZ\ signal events, extracted from the invariant
mass spectra in the data and corrected for selection efficiency and
acceptance effects. These are extracted from reconstructed signal
Monte Carlo events that pass the same selection criteria as events in
data. The efficiency corrected yields are then used to calculate the
ratios of branching fractions relative to the \Omegapi\ mode, yielding 
\begin{align*}
  \frac{{\cal B}(\Omegapipipi)}{{\cal B}(\Omegapi)} 
         &= \ROmegapipipi \pm \RstatOmegapipipi\mathrm{(stat.)} 
         \pm \RsysOmegapipipi\mathrm{(syst.)},\\
  \frac{{\cal B}(\XiKpipi)}{{\cal B}(\Omegapi)}     
         &= \RXiKpipi \pm \RstatXiKpipi\mathrm{(stat.)} 
         \pm \RsysXiKpipi\mathrm{(sys.)}.
\end{align*}
The individual contributions to the systematic uncertainties are
discussed in detail in Section~\ref{sub:SysStudies}. 

\subsection{Systematic Studies}\label{sub:SysStudies}
Although the decay topologies are slightly different in the various
decay modes, the systematic uncertainties on the hyperon selection
efficiencies largely cancel in the ratios of branching
fractions.  
The following sources of systematic uncertainties are considered in
the measurement of the ratios of branching fractions and summarized in
Table~\ref{tab:Systematics}. 
\begin{itemize}
\item Monte Carlo Simulation:\par
    The statistical uncertainty on the selection efficiencies is taken
    into account in the systematic uncertainty.

    In multi-body decay modes, the decay can occur via
    intermediate resonances. Therefore, signal Monte Carlo
    samples containing the \FXiSKpi, the \FXiKstarzpi, and the
    \FXiSKstarz\ decay modes are generated. The difference in
    selection efficiencies relative to the uniform-phase-space sample is
    included as an  uncertainty on the selection efficiency. For the
    \FOmegapipipi\ mode, the systematic uncertainty associated with
    intermediate resonances is assumed to be the same as that for the
    topologically similar mode \FXiKpipi.   

    The $p^*$ spectrum of the \OmegacZ\ in data and Monte Carlo is
    slightly (but not significantly) different. The effect on the
    selection efficiency is calculated and a systematic uncertainty
    accounting for the difference is assigned. This contribution is
    correlated for the decay mode in the numerator and in the
    denominator and is treated accordingly in the calculation of the
    systematic uncertainty on the ratio of branching fractions. 
\begin{table}[!b]
  \centering\small
  \caption{Systematic uncertainties considered in the measurement of
    the ratios of branching fractions. The individual contributions
    are given, and added in quadrature to determine the total
    systematic uncertainty. Dashes indicate sources that are assumed
    to cancel in the ratio of branching fractions.}\vspace*{1ex}   
  \label{tab:Systematics}
\begin{tabular}{l|cc}\hline\hline
  \both ~ & $\frac{\displaystyle\up {\cal B}(\FOmegapipipi)}{\displaystyle\down {\cal B}(\FOmegapi)}$
  & $\frac{\displaystyle\up {\cal B}(\FXiKpipi)}{\displaystyle\down {\cal B}(\FOmegapi)}$ \\\hline
  \up   Monte Carlo Statistics       & \RMCMCOmegapipipi  & \RMCMCXiKpipi \\ 
        $p^\ast$ Reweighting         & \RMCPAOmegapipipi  & \RMCPAXiKpipi \\ 
  \down Resonance Structure          & \RMCRSOmegapipipi  & \RMCRSXiKpipi \\\hline
        Extraction of Signal Yield   & \RFMOmegapipipi    & \RFMXiKpipi    \\
        Particle ID \& Tracking      & \RPIDOmegapipipi   & \RPIDXiKpipi    \\
        \Omegam Branching Fraction   & \ROMOmegapipipi    & \ROMXiKpipi    \\
  \down Multiple Candidates          & \RMCOmegapipipi    & \RMCXiKpipi   \\\hline
  \up Total Systematic Uncertainty   & \RsysOmegapipipi   & \RsysXiKpipi   \\\hline\hline
\end{tabular}
\end{table}

\item Extraction of the Signal Yield:\par
  In the fit to the invariant mass spectra, the description of the
  background shape is varied for the purpose of estimating the
  systematic uncertainty. In addition, the width of the widest
  Gaussian in the description of the signal shape is varied by a
  factor of two, and the fit range is varied from $2.6<m<2.8\gevcc$ to
  $2.4<m<3.0\gevcc$. The observed variations in the signal yield are 
  added as a systematic uncertainty. 
\item Particle Identification and Tracking:\par
  The number of primary mesons is different for some of the modes. A
  systematic uncertainty of $1\%$ is added 
  for each additional primary pion, estimated from the uncertainty on
  the efficiency of the pion identification. A systematic
  uncertainty of $1\%$ is assigned to account for different kaon
  identification algorithms used for the primary and non-primary
  kaons. 

  In order to account for the difference in tracking efficiency in data
  and Monte Carlo, a correction of $0.25\%$ with a systematic
  uncertainty of  $1.4\%$ is applied per track. In the ratio of
  branching fractions this amounts to a $0.5\%$ correction to the
  selection efficiency with a systematic uncertainty of $2.8\%$.

\item Branching Fraction:\par
  The \Omegam\ is identified in its decay to \LambdaZ\Km, which has a
  branching fraction of $(67.8\pm 0.7)\%$. This uncertainty in the
  branching fraction is added to the systematic uncertainty for the
  ratio involving the \Xim\ only. The uncertainty in the
  \LambdaZ\ branching fraction to $p\pi^-$ cancels in the ratio of
  branching fractions for all modes.   

\item Multiple Candidates:\par
  A possible source of background comes from the presence of multiple
  \OmegacZ\ candidates in an event, which share one or more
  tracks. Predominantly, the same hyperon combines with one or more
  primary tracks to form such multiple candidates. In general, these
  candidates are distributed over a large mass range. However, in
  cases where these candidates share tracks, their masses might be
  correlated, which could lead to a pile-up of candidates in the
  signal region. The mass distribution of \OmegacZ\ in events with
  multiple candidates is studied in data as well as in Monte Carlo signal
  samples. We select events in which one of the candidates falls
  inside a $\pm 3\ \mathrm{HWHM}$ mass window around the nominal peak
  position.  In data, two, four, and four events with such multiple
  candidates are observed in the \FOmegapi, \FOmegapipipi, and
  \FXiKpipi, respectively, where the second candidate also lies in the
  signal region defined above. From studies of these candidates in a
  larger mass window, these multiple candidates are
  observed to be evenly distributed in mass for all modes. Therefore,
  they form part of the background when the mass spectra are fit with
  a polynomial. This indicates that there is no statistically
  significant peaking under the signal peak in data. 

  The relative sizes of the peaks from incorrectly reconstructed
  multiple candidates and true signal candidates in Monte
  Carlo are $0.2\%$, $1.2\%$ and $1.3\%$ in the \FOmegapi,
  \FOmegapipipi, and \FXiKpipi\ mode, respectively. This fraction is
  assigned as a systematic uncertainty. 

\item Other Sources of Peaking Background:\par
  Possible sources of peaking backgrounds are studied with continuum
  Monte Carlo samples. The number of peaking background events
  observed is consistent with zero and therefore no systematic
  uncertainty is assigned.  
\end{itemize}

The individual sources of systematic uncertainties are added in
quadrature. The total uncertainty on each ratio is given in
Table~\ref{tab:Systematics}. 

\subsection{Limit Calculation for the \Omegapipipi Mode}
No significant excess of signal events over the background is
observed in the \FOmegapipipi\ mode. Therefore, a limit at the $90\%$
confidence level (CL) on the ratio of branching fractions is
calculated. The limit is obtained from a Monte Carlo calculation for the
individual modes, using the measured yields and the statistical and
systematic uncertainties as inputs. All uncertainties are assumed to
be Gaussian.   
Integrating the positive part of the distribution of ratios of
branching fractions obtained from these Monte Carlo experiments, an
upper limit at the $90\%$ confidence level of 
\begin{align*}
  \frac{\displaystyle {\cal B}(\Omegapipipi)}
       {\displaystyle {\cal B}(\Omegapi)}     
       < \LOmegapipipi\qquad (90\%\mathrm{CL})
\end{align*}
is obtained.

\subsection{Cross-checks}
Cross-checks for the measurements of the ratios of branching fractions
are performed to verify the stability of the result.
\begin{itemize}
\item Charge asymmetry:\par
  To obtain the main result, no distinction is made between decays
  of the \OmegacZ\ and the anti-baryon \AOmegacZ. However,
  the selection efficiencies might be different due to
  differences in the interaction in material of particles and
  anti-particles. Particle and anti-particle candidates are selected
  and studied separately. The observed difference in
  efficiency-corrected yields for mode and anti-mode is largest in the
  \OmegacZ\ decay into \FXiKpipi, with a difference of $1.6\sigma$. All
  ratios of branching fractions are found to be consistent within the
  statistical uncertainties. 

\item Choice of the $p^\ast$ range:\par
  The main results for the ratios of branching fractions are obtained
  for a minimum $p^\ast$ of $2.8\gevc$. The ratios of 
  branching fractions, however, are expected to be independent of that
  choice. As a cross-check, these ratios are recalculated, requiring
  a $p^\ast$ in the range $2.6\gevc<p^\ast<3.2\gevc$ and
  $p^\ast>3.2\gevc$. All results are found to be consistent within
  statistical uncertainties.  
\end{itemize}

\section{SUMMARY}
\label{sec:Summary}
Data recorded with the \babar\ detector are analyzed to study
\OmegacZ\ production and decays. The \OmegacZ\ is reconstructed through
its decays into \FOmegapi and \FOmegapipipi (using $225$\invfb of
data), and into \FXiKpipi\ (using $230$\invfb). 

Based on the momentum spectrum of the \OmegacZ\ in the \epem\
rest frame in the full data set and in off-peak data only, the first 
observation of \OmegacZ\ production from $B$ decays is reported.

In the \FOmegapi\ decay mode, the \OmegacZ\ is observed with a
statistical significance of over $17\sigma$. This constitutes the
first observation of the \OmegacZ\ baryon above the $5\sigma$ level.  

From the observed signal yields, corrected for efficiency and
acceptance, the ratios of branching fractions relative to \Omegapi\
are measured to be 
\begin{align*}
  \frac{{\cal B}(\Omegapipipi)}{{\cal B}(\Omegapi)} 
         &= \ROmegapipipi \pm \RstatOmegapipipi\mathrm{(stat.)} 
         \pm \RsysOmegapipipi\mathrm{(syst.)},\\
  \frac{{\cal B}(\XiKpipi)}{{\cal B}(\Omegapi)}     
         &= \RXiKpipi \pm \RstatXiKpipi\mathrm{(stat.)} 
         \pm \RsysXiKpipi\mathrm{(sys.)},
\intertext{where the first uncertainty is statistical and the second is
         systematic. Due to the limited statistical significance for
         the decay mode \Omegapipipi, an upper limit on the ratio of
         branching fractions,}
  \frac{{\cal B}(\Omegapipipi)}{{\cal B}(\Omegapi)} &<
  \LOmegapipipi \qquad\mathrm{(90\% CL), }
\end{align*}
at the $90\%$ confidence level is set. All results mark a
considerable improvement, both in statistical and systematic
uncertainties, over the current world averages~\cite{bib:PDG2004}.  

\section{ACKNOWLEDGMENTS}
\label{sec:Acknowledgments}


\input include/acknowledgements

\end{document}

%% file: include/phys.tex
\def\LambdaZ     {\ensuremath{\Lambda}}
\def\OmegacZ     {\ensuremath{\Omega_\c^0}\xspace}
\def\AOmegacZ     {\ensuremath{\overline{\Omega}_\c^0}\xspace}

\def\XiS     {\ensuremath{\Xi^\ast(1530)^0}\xspace}

\def\Xim     {\ensuremath{\Xi^-}\xspace}

\def\Omegam     {\ensuremath{\Omega^-}\xspace}

\def\Omegapi      {\ensuremath{\OmegacZ\to\Omegam\pip}\xspace}

\def\Omegapipipi  {\ensuremath{\OmegacZ\to\Omegam\pip\pim\pip}\xspace}

\def\SigmaKKpi    {\ensuremath{\OmegacZ\to\Sigma^+\Km\Km\pip}\xspace}
\def\XiKpipi      {\ensuremath{\OmegacZ\to\Xim\Km\pip\pip}\xspace}

\def\FOmegapi      {\ensuremath{\Omegam\pip}\xspace}

\def\FOmegapipipi  {\ensuremath{\Omegam\pip\pim\pip}\xspace}

\def\FXiKpipi      {\ensuremath{\Xim\Km\pip\pip}\xspace}

\def\FXiSKstarz    {\ensuremath{\XiS\Kstarz}\xspace}
\def\FXiKstarzpi   {\ensuremath{\Xim\Kstarz\pip}\xspace}
\def\FXiSKpi       {\ensuremath{\XiS\Km\pip}\xspace}

%% file: include/results.tex
\def\EffOmegapi{\ensuremath{8.35}}
\def\DEffOmegapi{\ensuremath{0.07}}

\def\EffOmegapipipi{\ensuremath{4.41}}
\def\DEffOmegapipipi{\ensuremath{0.09}}
\def\EffXiKpipi{\ensuremath{5.63}}
\def\DEffXiKpipi{\ensuremath{0.10}}

\def\RYOmegapi{\ensuremath{138.5}}
\def\RYstatOmegapi{\ensuremath{14.8}}

\def\RYOmegapipipi{\ensuremath{11.8}}
\def\RYstatOmegapipipi{\ensuremath{7.5}}

\def\RYXiKpipi{\ensuremath{29.9}}
\def\RYstatXiKpipi{\ensuremath{13.6}}

\def\ROmegapipipi{\ensuremath{0.16}}
\def\RstatOmegapipipi{\ensuremath{0.10}}
\def\RMCMCOmegapipipi{\ensuremath{0.004}}
\def\RMCRSOmegapipipi{\ensuremath{0.003}}
\def\RMCPAOmegapipipi{\ensuremath{0.001}}
\def\RFMOmegapipipi{\ensuremath{0.028}}

\def\RPIDOmegapipipi{\ensuremath{0.006}}
\def\ROMOmegapipipi{\ensuremath{--}}
\def\RMCOmegapipipi{\ensuremath{0.002}}

\def\RsysOmegapipipi{\ensuremath{0.03}}

\def\RXiKpipi{\ensuremath{0.31}}
\def\RstatXiKpipi{\ensuremath{0.15}}
\def\RMCMCXiKpipi{\ensuremath{0.006}}
\def\RMCRSXiKpipi{\ensuremath{0.005}}
\def\RMCPAXiKpipi{\ensuremath{0.004}}
\def\RFMXiKpipi{\ensuremath{0.032}}

\def\RPIDXiKpipi{\ensuremath{0.011}}
\def\ROMXiKpipi{\ensuremath{0.003}}
\def\RMCXiKpipi{\ensuremath{0.004}}

\def\RsysXiKpipi{\ensuremath{0.04}}
 
\def\LOmegapipipi{\ensuremath{0.30}}

%% file: include/authors_conf05010.tex
\begin{center}
\small

The \babar\ Collaboration,
\bigskip

B.~Aubert,
R.~Barate,
D.~Boutigny,
F.~Couderc,
Y.~Karyotakis,
J.~P.~Lees,
V.~Poireau,
V.~Tisserand,
A.~Zghiche
\inst{Laboratoire de Physique des Particules, F-74941 Annecy-le-Vieux, France }
E.~Grauges
\inst{IFAE, Universitat Autonoma de Barcelona, E-08193 Bellaterra, Barcelona, Spain }
% ...

\end{center}\newpage

%% file: include/acknowledgements.tex
We are grateful for the 
extraordinary contributions of our \pep2\ colleagues in
achieving the excellent luminosity and machine conditions
that have made this work possible.
The success of this project also relies critically on the 
expertise and dedication of the computing organizations that 
support \babar.
The collaborating institutions wish to thank 
SLAC for its support and the kind hospitality extended to them. 
This work is supported by the
US Department of Energy
and National Science Foundation, the
Natural Sciences and Engineering Research Council (Canada),
Institute of High Energy Physics (China), the
Commissariat \`a l'Energie Atomique and
Institut National de Physique Nucl\'eaire et de Physique des Particules
(France), the
Bundesministerium f\"ur Bildung und Forschung and
Deutsche Forschungsgemeinschaft
(Germany), the
Istituto Nazionale di Fisica Nucleare (Italy),
the Foundation for Fundamental Research on Matter (The Netherlands),
the Research Council of Norway, the
Ministry of Science and Technology of the Russian Federation, and the
Particle Physics and Astronomy Research Council (United Kingdom). 
Individuals have received support from 
CONACyT (Mexico),
the A. P. Sloan Foundation, 
the Research Corporation,
and the Alexander von Humboldt Foundation.